\documentclass[oneside,twocolumn,showpacs,preprintnumbers,amsmath,amssymb]{revtex4}

\usepackage{amsmath}
\usepackage{amsfonts}
\usepackage{amssymb, multirow}
\usepackage{pdfsync}
\usepackage{epsfig}
\usepackage{graphicx}
\usepackage{enumitem}
\usepackage{array}
\usepackage{cleveref}
\usepackage[utf8]{inputenc}
\crefname{section}{§}{§§}

\begin{document}

\title{CSO$_{c}\,$ superpotentials\\[2mm]}

\author{Adolfo Guarino}

\affiliation{
\vspace{5mm}
Nikhef, Science Park 105, 1098 XG Amsterdam, The Netherlands
}

\begin{abstract}

Motivated by their applications to holographic RG flows and hairy black holes in Einstein-scalar systems, we present a collection of superpotentials driving the dynamics of $\,\mathcal{N}=2$ and $\,\mathcal{N}=1\,$ four-dimensional supergravities. These theories arise as consistent truncations of the electric/magnetic families of $\,\textrm{CSO}(p,q,r)_{c}\,$ maximal supergravities, with $\,{p+q+r=8}$, discovered by Dall'Agata et al. The $\,\mathcal{N}=2\,$ and $\,\mathcal{N}=1\,$ truncations describe $\,\textrm{SU}(3)\,$ and $\,\mathbb{Z}_{2} \times \textrm{SO}(3)\,$ invariant sectors, respectively,  and contain AdS$_{4}$ solutions preserving $\,\mathcal{N}=1,2,3,4\,$ supersymmetry within the full theories, as well as various gauge symmetries. Realisations in terms of non-geometric type IIB as well as geometric massive type IIA backgrounds are also discussed. The aim of this note is to provide easy to handle superpotentials that facilitate the study of gravitational and gauge aspects of the $\textrm{CSO}(p,q,r)_{c}$ maximal supergravities avoiding the technicalities required in their construction.

\end{abstract}

\pacs{04.65.+e, 11.25.Mj \\[1mm] NIKHEF 2015-028 \\[1mm] e-mail: aguarino@nikhef.nl}

\maketitle

\section{Introduction}
\label{Sec:Intro}

The dynamics of supergravity theories in four dimensions (4D) is conjectured to describe three-dimensional field theories (at large $N$) in a strongly coupled regime via the gauge/gravity duality \cite{Maldacena:1997re}. The reliability of the correspondence increases with the amount of supersymmetry, thus selecting the maximal $\,\mathcal{N}=8\,$ gauged supergravities in 4D \cite{Cremmer:1979up,deWit:1981eq,deWit:1982ig,deWit:2007mt} as preferred models where to test the duality \cite{Schwarz:2004yj,Bagger:2006sk,Bagger:2007jr,Gustavsson:2007vu,Aharony:2008ug}. However, highly supersymmetric models come along with a large number of fields filling the corresponding supermultiplets. For instance, the bosonic content of the $\,\mathcal{N}=8\,$ supergravity multiplet consists of the metric, $28$ vector fields spanning a gauge group $\textrm{G}$ (a.k.a. gauging) and $70$ scalar fields parameterising an $\textrm{E}_{7(7)}/\textrm{SU}(8)$ coset space \cite{Cremmer:1979up,Cremmer:1997ct}. Having such a large number of vectors and scalars makes the dynamics of maximal supergravities difficult to be analysed in full generality and it is at this point where the notion of consistent truncation comes to rescue. 

A consistent truncation of the maximal supergravity field content implies turning off most of the fields of the theory and retaining only a small but more tractable subset of it. The truncation is called consistent if, provided a given solution to the equations of motion of the truncated theory, then those of the full theory are also satisfied. In other words, solutions of the truncated theory correspond to solutions of the full theory in which the truncated fields have been set to zero. One way of performing a consistent truncation is to keep only the subset of fields which are invariant (singlets) under the action of a compact subgroup $\textrm{G}_{0} \subset \textrm{G}$ of the gauge group $\textrm{G}$ spanned by the vectors in the full theory \cite{Warner:1983vz}. Consistency requires the gauge group $\textrm{G}$ to be a subgroup of the U-duality group $\textrm{E}_{7(7)}$ of the maximal 4D \mbox{supergravities} \cite{Cremmer:1978km,deWit:2007mt} which is in turn embedded into the $\textrm{Sp}(56,\mathbb{R})$ symplectic group of electric/magnetic transformations of the theory \cite{deWit:2002vt}. Schematically,
\begin{equation}
\textrm{G}_{0} \, \subset \, \textrm{G}  \, \subset \, \textrm{E}_{7(7)} \, \subset \, \textrm{Sp}(56,\mathbb{R}) \ .
\end{equation}

As first noticed in the vacua classification of \cite{DallAgata:2011aa} and then made more precise in \cite{Dall'Agata:2012bb,Dall'Agata:2014ita}, the embedding of the vector fields spanning $\,\textrm{G}\,$ into the electromagnetic group $\,\textrm{Sp}(56,\mathbb{R})\,$ may allow for certain freedom, or symplectic deformation. In ref.~\cite{Dall'Agata:2012bb}, a one-parameter family of $\textrm{G}=\textrm{SO}(8)$ gauged supergravities generalising that of de Wit and Nicolai \cite{deWit:1981eq,deWit:1982ig} was built and the structure of critical points of the associated scalar potential partially explored by looking at the $\textrm{G}_{0} = \textrm{G}_{2}$ consistent truncation retaining the metric and two real scalars. The \mbox{electric/magnetic} deformation parameter was denoted $c$ and the family of new theories dubbed $\textrm{SO}(8)_{c}$. Soon after, the scalar potential associated to different truncations of the SO(8)$_{c}$ theories to $\textrm{G}_{0} = \textrm{SU}(3)$ \cite{Borghese:2012zs} and $\textrm{G}_{0} = \textrm{SO}(4)$ \cite{Borghese:2013dja} invariant sectors were put forward and their structure of critical points unraveled. In all the cases, the number of critical points, their location in field space and the corresponding value of the scalar potential $V_{0}$ turned out to change as a function of the electromagnetic or symplectic parameter $c$. 

These remarkable facts immediately rose questions about the possible embeddings of the deformed theories (and their novel critical points) into higher-dimensional theories as well as about their conjectured gauge duals. Trying to answer these questions requires a combined understanding of: $i)$~The embedding tensor formalism \cite{deWit:2007mt} in order to derive simple supergravity models based on electric/magnetic gaugings \cite{Dall'Agata:2014ita}. $ii)$~Techniques of dimensional reduction/oxidation of supergravity theories (or exceptional versions thereof \cite{Hohm:2013pua,Hohm:2013uia}). $iii)$~Three-dimensional field theories and their connection to the theory of M2/D2-branes \cite{Schwarz:2004yj,Bagger:2006sk,Bagger:2007jr,Gustavsson:2007vu,Aharony:2008ug}. These three aspects have recently been clarified for a cousin of the SO(8)$_{c}$ theories, namely, the electric/magnetic family of ISO(7)$_{c}$ maximal supergravities \cite{Guarino:2015jca,Guarino:2015qaa}.

This note aims to contribute to the first ``vertex of the triangle". We will present a collection of superpotentials controlling the scalar dynamics in $\,{\mathcal{N}=2}\,$ and $\mathcal{N}=1$ consistent truncations of $\,\textrm{CSO}_{c} \equiv\textrm{CSO}(p,q,r)_{c}$ maximal supergravities, with $p+q+r=8$, based on $\textrm{G}_{0}=\textrm{SU}(3)\,$ and $\,\textrm{G}_{0}=\mathbb{Z}_{2} \times \textrm{SO}(3)\,$ invariant sectors. Despite their simplicity, the truncated theories turn out to capture a broad set of the critical points studied in the undeformed $(c=0)$ theories, their electric/magnetic deformations and more. By the end of the note, we will briefly touch on (non-)geometric string/M-theory incarnations of the $\textrm{CSO}_{c}$ maximal supergravities.

\section{Electric/magnetic deformations}
\label{Sec:Elec/Mag}

We start by introducing the families of $\textrm{CSO}(p,q,r)_{c}$ maximal supergravities, with $p+q+r=8$, developed by Dall'Agata et al in \cite{DallAgata:2011aa,Dall'Agata:2012bb,Dall'Agata:2014ita}. They correspond to symplectic deformations of the $\textrm{CSO}(p,q,r)$ theories of Hull \cite{Hull:1984yy,Hull:1984vg,Hull:1984qz,Hull:1984rt} (see also \cite{Cordaro:1998tx}) by turning on an electric/magnetic deformation parameter $c$. In \cite{Dall'Agata:2014ita}, it was shown that only the semisimple $\,\textrm{SO}(p,q) \equiv \textrm{CSO}(p,q,0)\,$ as well as the non-semisimple $\,\textrm{ISO}(p,q) \equiv \textrm{CSO}(p,q,1)\,$ gaugings turn out to admit such a deformation. Remarkably, it turns out to be a discrete (on/off) deformation for the latter \cite{Dall'Agata:2014ita}. The undeformed theories are then called ``electric" and are recovered at $c=0$. Turning on the deformation $c\,$ modifies the combinations of electric and magnetic vector fields $A^{\mathbb{M}}_{\mu}=(A^{[AB]}_{\mu},A_{\mu \, [AB]})$ which are to enter the gauge covariant derivatives $D_{\mu}$ of the maximal supergravity theory. We have introduced a fundamental E$_{7(7)}$ index $\mathbb{M}=1,...,56$ as well as a fundamental SL(8) index $A=1,...,8$. Then, the $\textrm{E}_{7(7)} \supset \textrm{SL}(8) $ decomposition $\textbf{56} \rightarrow \textbf{28'} + \textbf{28}$ simply reflects the electric (28 of them) and magnetic (28 of them) nature of the vector fields in maximal supergravity. In the undeformed case $(c = 0)$, only the electric vectors $A^{[AB]}_{\mu}$ enter the covariant derivative. After turning  on the symplectic deformation, this changes to   
\begin{equation}
\label{D_cov}
\begin{array}{lll}
D_{\mu} & = & \partial_{\mu}  - g \,  X_{\mathbb{M}} \, A^{\mathbb{M}} _{\mu}  \\[2mm]
              & = & \partial_{\mu}  - g \, \left(  X_{[AB]} \, A^{[AB]} _{\mu} + X^{[AB]} \, A_{\mu \, [AB]} \right) \ ,
\end{array}
\end{equation}
containing both electric and magnetic charges ${X_{\mathbb{M}}=(X_{[AB]} , X^{[AB]})}$, the latter being proportional to the deformation parameter $c$. All the charges are specified by an \textit{embedding tensor} ${\Theta_{\mathbb{M}}}^{\alpha}$ upon contraction with the $\textrm{E}_{7(7)}$ generators $t_{\alpha}$  
\begin{equation}
\label{charges}
X_{\mathbb{M}} = {\Theta_{\mathbb{M}}}^{\alpha} \, t_{\alpha}
\hspace{5mm} \textrm{ with } \hspace{5mm}
{X_{\mathbb{MN}}}^{\mathbb{P}} = {\Theta_{\mathbb{M}}}^{\alpha} \, {[t_{\alpha}]_{\mathbb{N}}}^{\mathbb{P}} \ ,
\end{equation}
where $\alpha=1,...,133$ is an adjoint index of $\textrm{E}_{7(7)}$. Therefore, the embedding tensor ${\Theta_{\mathbb{M}}}^{\alpha}$ selects which electric/magnetic combinations of vectors are to span the gauge group $\textrm{G} \subset \textrm{E}_{7(7)}$ of the maximal supergravity. In order to guarantee that only $\,28\,$ linearly independent combinations actually enter the gauging, a set of quadratic constraints of the form
\begin{equation}
\label{Orthog_cond}
\Omega^{\mathbb{MN}} \, {\Theta_{\mathbb{M}}}^{\alpha} \, {\Theta_{\mathbb{N}}}^{\beta} = 0 
\hspace{2mm} \textrm{ with } \hspace{2mm}
\Omega^{\mathbb{MN}} = \left(\begin{array}{ll} 0_{28} & \mathbb{I}_{28} \\ -\mathbb{I}_{28} & 0_{28}  \end{array}\right) \ ,
\end{equation}
has to be satisfied by ${\Theta_{\mathbb{M}}}^{\alpha}$. In this sense, eq.~(\ref{Orthog_cond}) can be viewed as an orthogonality condition for the charges \cite{deWit:2007mt}.

Let us first look at the family of $\textrm{SO}(8)_{c}$ maximal supergravities \cite{Dall'Agata:2012bb} that leaves invariant the metric 
\begin{equation}
\eta = \textrm{diag}(1,1,1,1,1,1,1, 1) \ .
\end{equation}
In this case, the embedding tensor ${\Theta_{\mathbb{M}}}^{\alpha}$ takes the simple form
\begin{equation}
\label{Theta_comp8}
\begin{array}{lllll}
{\Theta_{[AB]}}^{CD} = \delta_{AB}^{CD}
&\hspace{2mm} , \hspace{2mm}&
\Theta^{[AB] \, CD} = -  c  \, \delta_{AB}^{CD} & .
\end{array}
\end{equation}
where the index $\alpha$ in the ${\Theta_{\mathbb{M}}}^{\alpha}$ components in (\ref{Theta_comp8}) runs over the linear combinations
\begin{equation}
T_{CD} \equiv {t_{C}}^{D} - {t_{D}}^{C} \ ,
\end{equation}
of $\textrm{SL}(8)$ generators ${t_{C}}^{D}$ in the $\textrm{SL}(8)$ decomposition of $\textrm{E}_{7(7)}$ (see Appendix). These are precisely the 28 generators of SO(8).  Plugging (\ref{Theta_comp8}) into the $X_{\mathbb{M}}$ charges of (\ref{charges}) and substituting the result in (\ref{D_cov}) gives rise to a covariant derivative of the form
\begin{equation}
\label{D_cov_SO8}
D_{\mu} = \partial_{\mu} - g \Big(  A^{[CD]}_{\mu} -  c \, A_{\mu \, [CD]} \Big) \ .
\end{equation}
From (\ref{D_cov_SO8}) one sees that the parameter $c$ sets the linear combinations of electric and magnetic vectors that enter the gauging. The electric SO(8) gauged supergravity of \cite{deWit:1981eq,deWit:1982ig} is recovered at $\,c=0\,$.

The $\textrm{SO}(7,1)_{c}$ and $\textrm{ISO}(7)_{c}\equiv\textrm{CSO}(7,0,1)_{c}$ gaugings can be jointly described if applying the index splitting $A=(m,8)$ with ${m=1,..,7}$. The invariant metrics preserved by the gaugings can be written as 
\begin{equation}
\eta = \textrm{diag}(1,1,1,1,1,1,1, \epsilon_{1}  \epsilon_{2}) \ ,
\end{equation}
with $(\epsilon_{1},\epsilon_{2})$ being $(1,-1)$ and $(0,1)$ for the SO(7,1)$_{c}$ and ISO(7)$_{c}$ gaugings, respectively.
The embedding tensor ${\Theta_{\mathbb{M}}}^{\alpha}$ is then given by
\begin{equation}
\label{Theta_comp7}
\begin{array}{lllll}
{\Theta_{[mn]}}^{pq} = \delta_{mn}^{pq}
&\hspace{2mm} , \hspace{2mm}&
\Theta^{[mn] \, pq} = - \epsilon_{1} \, c  \, \delta_{mn}^{pq} & , \\[2mm]
{\Theta_{[m8]}}^{p8} = \delta_{m}^{p}
&\hspace{2mm} , \hspace{2mm}&
\Theta^{[m8] \, p8} = - \epsilon_{2} \, c  \, \delta_{m}^{p} & ,
\end{array}
\end{equation}
where the index $\alpha$ in the ${\Theta_{\mathbb{M}}}^{\alpha}$ components in (\ref{Theta_comp7}) runs over the linear combinations
\begin{equation}
T_{pq} \equiv {t_{p}}^{q} - {t_{q}}^{p}
\hspace{4mm} \textrm{ and } \hspace{4mm}
T_{p8} \equiv -\epsilon_{1}  \, {t_{p}}^{8} - {t_{8}}^{p} 
\end{equation}
of $\textrm{SL}(8)$ generators ${t_{A}}^{B}$. There is then an SO(7) subgroup spanned by $\,T_{pq}\,$ which is extended to either SO(7,1) or ISO(7) by the seven generators $\,T_{p8}$. \mbox{Plugging} (\ref{Theta_comp7}) into (\ref{charges}) and substituting again in (\ref{D_cov}) gives rise to a covariant derivative of the form
\begin{equation}
\label{D_cov_SO7}
D_{\mu} = \partial_{\mu} - g \Big(  A^{[pq]}_{\mu} - \epsilon_{1} c \, A_{\mu \, [pq]} \Big) - g \Big(  A^{[p8]}_{\mu} - \epsilon_{2} c \, A_{\mu \, [p8]}\Big) .
\end{equation}
As noticed in \cite{Dall'Agata:2014ita}, taking $\,c\neq 0\,$ in (\ref{D_cov_SO7}) translates into all the generators being gauged dyonically in the SO(7,1)$_{c}$ case whereas only the seven flat generators $T_{p8}$ are gauged dyonically in the ISO(7)$_{c}$ case with $\,\epsilon_{1}=0\,$.

The SO(6,2)$_c$ and $\textrm{ISO}(6,1)_c \equiv \textrm{CSO}(6,1,1)_{c}$ gaugings can be jointly analysed in a similar manner. This time we split the fundamental SL(8) index as $A=(1,a,8)$ with $a=2,...,7$. The invariant metrics preserved by the gaugings are now given by
\begin{equation}
\eta = \textrm{diag}(-1,1,1,1,1,1,1, \epsilon_{1} \epsilon_{2}) \ ,
\end{equation}
with $(\epsilon_{1},\epsilon_{2})$ being $(1,-1)$ and $(0,1)$ for the SO(6,2)$_c$ and ISO(6,1)$_c$ gaugings, respectively. The embedding tensor ${\Theta_{\mathbb{M}}}^{\alpha}$ has components
\begin{equation}
\label{Theta_comp6}
\begin{array}{lllll}
{\Theta_{[ab]}}^{cd} = \delta_{ab}^{cd}
&\hspace{2mm} , \hspace{2mm}&
\Theta^{[ab] \, cd} = - \epsilon_{1} \, c  \, \delta_{ab}^{cd} & , \\[2mm]
{\Theta_{[18]}}^{18} = -1
&\hspace{2mm} , \hspace{2mm}&
\Theta^{[18] \, 18} =  -\epsilon_{2} \, c   & , \\[2mm]
{\Theta_{[1b]}}^{1d} = -\delta_{b}^{d}
&\hspace{2mm} , \hspace{2mm}&
\Theta^{[1b] \, 1d} =  - \epsilon_{1} \, c  \, \delta_{b}^{d} & , \\[2mm]
{\Theta_{[a8]}}^{c8} = \delta_{a}^{c}
&\hspace{2mm} , \hspace{2mm}&
\Theta^{[a8] \, c8} =  - \epsilon_{2} \, c  \, \delta_{a}^{c} & ,
\end{array}
\end{equation}
with the index $\alpha$ in ${\Theta_{\mathbb{M}}}^{\alpha}$ running this time over the linear combinations
\begin{equation}
\begin{array}{lllll}
T_{cd} \equiv {t_{c}}^{d} - {t_{d}}^{c}
&\hspace{4mm} , \hspace{4mm} &
T_{18} \equiv \epsilon_{1} \, {t_{1}}^{8} - {t_{8}}^{1} & ,\\[2mm]
T_{1d} \equiv - {t_{1}}^{d} - {t_{d}}^{1}
& \hspace{4mm} , \hspace{4mm}&
T_{c8} \equiv - \epsilon_{1} \, {t_{c}}^{8} - {t_{8}}^{c} & ,
\end{array}
\end{equation}
of $\textrm{SL}(8)$ generators ${t_{A}}^{B}$. The covariant derivative in this case takes the form
\begin{equation}
\label{D_cov_SO6}
\begin{array}{lll}
D_{\mu} &=& \partial_{\mu} \\
&-&  g \Big(  A^{[cd]}_{\mu} -  \epsilon_{1}  c \, A_{\mu \, [cd]} \Big) + g \Big(  A^{[18]}_{\mu} + \epsilon_{2} c \, A_{\mu \, [18]}\Big) \\[2mm]
 & + & g \Big(  A^{[1d]}_{\mu} + \epsilon_{1} c \, A_{\mu \, [1d]} \Big) - g \Big(  A^{[c8]}_{\mu} - \epsilon_{2} c \, A_{\mu \, [c8]} \Big) \ .
\end{array}
\end{equation}
Taking again $c \neq 0$ in (\ref{D_cov_SO6}), all the generators are gauged dyonically in the SO(6,2)$_c$ case. For the ISO(6,1)$_c$ gaugings, only the seven flat generators $T_{18}$ and $T_{c8}$ are gauged dyonically as $\epsilon_{1}=0$, similar to what happened in the ISO(7)$_{c}$ case.

The rest of $\textrm{CSO}(p,q,r)$ gaugings, with ${p+q+r=8}$, that admit symplectic deformations are the families of $\textrm{SO}(5,3)_{c}\,$ and $\,\textrm{ISO}(5,2)_{c}\equiv \textrm{CSO}(5,2,1)_{c}\,$ gaugings leaving invariant the metrics
\begin{equation}
\label{eta_SO52}
\eta = \textrm{diag}(1,-1,1,-1,1,\epsilon_{1} \epsilon_{2},1, 1) \ ,
\end{equation}
with $(\epsilon_{1},\epsilon_{2})$ being $(1,-1)$ and $(0,1)$ respectively, as well as the $\,\textrm{SO}(4,4)_{c}\,$ and $\,\textrm{ISO}(4,3)_{c}\equiv \textrm{CSO}(4,3,1)_{c}\,$ ones with invariant metrics
\begin{equation}
\eta = \textrm{diag}(1,-1,1,-1,1,-1, 1 , \epsilon_{1} \epsilon_{2}) \ ,
\end{equation}
where $(\epsilon_{1},\epsilon_{2})$ are respectively given by $(1,-1)$ and $(0,1)$. The derivation of the corresponding embedding tensors and covariant derivatives proceeds as for the previous cases without surprises. As before, only the seven flat generators are gauged dyonically for the ISO(5,2)$_{c}$ and ISO(4,3)$_{c}$ gaugings. For the sake of brevity, we are not presenting the expressions here.

Apart from covariantising the derivatives in (\ref{D_cov}), \mbox{turning} on  a gauging drastically modifies the dynamics of the scalar fields in the theory by introducing a scalar potential \cite{deWit:2007mt}
\begin{equation}
\label{V_general}
V(\mathcal{M})  =  \dfrac{g^{2}}{672}  {X_{\mathbb{MN}}}^{\mathbb{R}}  {X_{\mathbb{PQ}}}^{\mathbb{S}} \mathcal{M}^{\mathbb{MP}} \big(  \mathcal{M}^{\mathbb{NQ}}  \mathcal{M}_{\mathbb{RS}} +   7 \,  \delta^{\mathbb{Q}}_{\mathbb{R}} \, \delta^{\mathbb{N}}_{\mathbb{S}}  \big) \ .
\end{equation}
In the above formula, the 70 scalars of maximal supergravity are encoded into a coset representative ${\mathcal{V} \in \textrm{E}_{7(7)}/\textrm{SU}(8)}$ which transforms under global $\textrm{E}_{7(7)}$ transformations from the left and local SU(8) ones from the right. This coset representative is then used to build the scalar-dependent matrix $\mathcal{M}_{\mathbb{MN}}$ as $\mathcal{M}=\mathcal{V} \, \mathcal{V}^{t}$, whose inverse $\mathcal{M}^{\mathbb{MN}}$ appears in (\ref{V_general}) together with the tensor ${X_{\mathbb{MN}}}^{\mathbb{P}}$ already introduced in (\ref{charges}). The kinetic terms for the scalars then follow from the standard coset constructions yielding a Einstein-scalar Lagrangian of the form
\begin{equation}
\label{L_scalars}
e^{-1} \mathcal{L}_{\textrm{E-s}} = \tfrac{1}{2} R + \tfrac{1}{96}  \textrm{Tr}\left( D_{\mu}\mathcal{M} \, D^{\mu}\mathcal{M}^{-1} \right) -  V(\mathcal{M}) \ .
\end{equation}
In this note we are setting all the vector fields to zero, so $D_{\mu} \rightarrow \partial_{\mu}$ in all the forthcoming formulas.

\section{$\mathcal{N}=2\,$ superpotentials}
\label{Sec:N2}

After shortly reviewing the electric/magnetic CSO$_{c}$ gaugings of maximal supergravity, we now move on towards our actual target: provide $\mathcal{N}=2$ truncations based on a $\textrm{G}_{0}=\textrm{SU}(3)$ invariant sector \cite{Warner:1983vz} that allow for an easy rewriting of the Lagrangian (\ref{L_scalars}).

The $\,\textrm{SO}(8)_{c}\,$, $\,\textrm{SO}(7,1)_{c}\,$, $\,\textrm{ISO}(7)_{c}\,$, $\,\textrm{SO}(6,2)_{c}\,$ and $\textrm{ISO}(6,1)_{c}\,$ gaugings, they all contain an $\textrm{SU}(3)$ subgroup within their maximal compact subgroups and, therefore, can accommodate such a truncation. The relevant chain of embeddings is given by
\begin{equation}
\label{N=2_chain}
\begin{array}{lcl}
& \textrm{SO}(6) & \\[1mm]
\textrm{SO}(8) \supset  \textrm{SO}(7)  \supset   
&  \textrm{or}& \supset  \textrm{SU}(3)  \ .  \\[1.5mm]
&  \textrm{G}_{2} & 
\end{array}
\end{equation} 
Truncating the $\,\mathcal{N}=8\,$ supergravity multiplet with respect to this SU(3) preserves $\mathcal{N}=2$ supersymmetry -- the $\textbf{8}$ gravitini of the maximal theory decompose as $\textbf{8} \rightarrow \textbf{1} + \textbf{1} + \textbf{3} + \textbf{3}$ under $\textrm{SU}(3) \subset \textrm{SU}(8)$, thus providing two singlets -- and retains the metric, two vector fields (we are setting to zero) and six real scalars. The scalars parameterise a scalar manifold ${\mathcal{M}_{\textrm{scal}} = \mathcal{M}_{\textrm{SK}} \times \mathcal{M}_{\textrm{QK}}}$ consisting of a special K\"ahler (SK) piece $\mathcal{M}_{\textrm{SK}}=\textrm{SU}(1,1)/\textrm{U}(1)$ and a quaternionic K\"ahler (QK) piece $\mathcal{M}_{\textrm{QK}}=\textrm{SU}(2,1)/\textrm{U}(2)$, accounting for one vector multiplet and one hypermultiplet.

The six real scalars in the truncation are associated to SU(3)-invariant combinations of E$_{7(7)}$ generators \cite{Guarino:2015qaa}. In the SL(8) basis, these are given by
\begin{equation}
\label{Gener_SU3}
\begin{array}{cclc}
g_{1} &=& {t_{3}}^{3} + {t_{5}}^{5} + {t_{7}}^{7} +  {t_{2}}^{2} + {t_{4}}^{4} + {t_{6}}^{6} - 3 \, ({t_{1}}^{1} + {t_{8}}^{8}) & , \\[2mm]

g_{2} &=&  {t_{8}}^{1} & , \\[2mm]

g_{3} &=&  {t_{1}}^{1} - {t_{8}}^{8} & , \\[2mm]

g_{4} &=&  t_{1238} + t_{1458} + t_{1678}  & ,  \\[2mm]

g_{5} &=&  t_{8357} - t_{8346} - t_{8256} - t_{8247} & ,  \\[2mm]

g_{6} &=&  t_{8246} - t_{8257} - t_{8347} - t_{8356}   & ,
\end{array}
\end{equation}
which are used to construct the coset representative ${\mathcal{V}=\mathcal{V}_{\textrm{SK}} \times \mathcal{V}_{\textrm{QK}}}$ upon the exponentiations
\begin{equation}
\label{coset_SU3}
\begin{array}{llll}
\mathcal{V}_{\textrm{SK}}  &=&  e^{-12 \, \chi_{1} \, g_{4}} \, e^{\frac{1}{4} \, \varphi_{1} \, g_{1}}& ,  \\[2mm]
\mathcal{V}_{\textrm{QK}}  &=&  e^{a \,  g_{2} \,-\, 6 \, (\zeta \, g_{5} \,+\, \tilde{\zeta} \,g_{6}) } \, e^{\phi \, g_{3}} & .
\end{array}
\end{equation}
With the above coset representative $\mathcal{V}$, the scalar-dependent matrix $\mathcal{M}$ entering (\ref{L_scalars}) is immediately obtained as $\,\mathcal{M}=\mathcal{V} \, \mathcal{V}^{t}$.  Equipped with the embedding tensors of the previous section for the set of CSO$_{c}$ gaugings compatible with $\textrm{G}_{0}=\textrm{SU}(3)$, it is a tedious exercise to work out the scalar Lagrangian. Plugging $\,\mathcal{M}\,$ into (\ref{L_scalars}) produces kinetic terms of the form
\begin{equation}
\label{LKin_c-map}
\begin{array}{lll}
e^{-1}  \mathcal{L}_{\textrm{kin}} &=& -\frac{3}{4} \left[ (\partial \varphi_{1} )^2 + e^{2 \varphi_{1}} \, (\partial \chi_{1} )^2\right]  \\[3mm]
&-&   (\partial \phi )^2  - \frac{1}{4} e^{2 \phi}  \left( (\partial \zeta)^{2}  +   (\partial \tilde{\zeta})^{2} \right) \\[3mm]
&-& \frac{1}{4} e^{4 \phi} \left( \partial a  +\frac{1}{2}\, (\zeta \, \partial \tilde{\zeta} - \tilde{\zeta}  \, \partial \zeta)  \right)^2 \ ,
\end{array}
\end{equation}
and a lengthy expression for the scalar potential $V$ in (\ref{V_general}) which depends on the particular choice of gauging. We will refrain from displaying the results here since, as stated in the abstract, we actually want to present them in a more concise $\mathcal{N}=2$ form.

To this end, we will start by rewriting the kinetic terms (\ref{LKin_c-map}) encoding the geometry of the scalar manifold $\mathcal{M}_{\textrm{scal}}$. The SK manifold spanned by the scalars $(\chi_{1},\varphi_{1})$ can be described in an $\mathcal{N}=2$ fashion by first complexifying to an upper-plane parameterisation $\Phi_{1}$ and then moving to a unit-disk parameterisation $z$ 
\begin{equation}
\label{SK_new variables}
(\chi_{1}, \varphi_{1}) \,\, \Rightarrow \,\, \Phi_{1} \equiv -\chi_{1} + i \, e^{-\varphi_{1}} \,\, \Rightarrow \,\, z \equiv \frac{\Phi_{1}-i}{\Phi_{1}+i} \ .
\end{equation}
In this way, the kinetic terms in (\ref{LKin_c-map}) for the scalars serving as coordinates in the SK manifold are expressed as
\begin{equation}
\label{LagSK}
e^{-1} \mathcal{L}^{\textrm{SK}}_{\textrm{kin}}  =  3 \, \frac{\partial \Phi_{1} \,  \partial \bar{\Phi}_{1} }{( \Phi_{1} -\bar{\Phi}_{1})^2}  = - 3 \, \frac{\partial z \,  \partial \bar{z}}{( 1- |z|^2)^2} \ .
\end{equation}
The geometry of the QK manifold is encoded in the kinetic terms for the four real scalars $(\phi,a,\zeta,\tilde{\zeta})$ in (\ref{LKin_c-map}). We can alternatively describe the geometry using two real $(\lambda , \sigma)$ and one complex $\psi$ fields \cite{Halmagyi:2011xh}
\begin{equation}
\label{QK_new variables_1}
\lambda \equiv e^{-2 \phi}
\hspace{3mm} , \hspace{3mm}
\sigma \equiv  a
\hspace{3mm} , \hspace{3mm}
\psi \equiv \tfrac{1}{2} (\tilde{\zeta} + i \, \zeta) \ , 
\end{equation}
in terms of which
\begin{equation}
\label{LagQK_1C}
\begin{array}{lll}
e^{-1} \mathcal{L}^{\textrm{QK}}_{\textrm{kin}}  &=& - \dfrac{1}{4 \lambda^2}  (\partial \lambda )^2  - \dfrac{1}{\lambda}   (\partial \psi ) (\partial \bar{\psi} ) \\[3mm]
&-& \dfrac{1}{4 \lambda^2} \Big( \partial \sigma  - i \, (\psi \, \partial \bar{\psi} - \bar{\psi}  \, \partial \psi)  \Big)^2 \ ,
\end{array}
\end{equation}
or using two complex fields $\,(\zeta_{1} , \zeta_{2})\,$ related to the previous ones by
\begin{equation}
\label{QK_new variables_2}
\lambda \equiv \dfrac{1 - |\zeta_{1}|^2 - |\zeta_{2}|^2}{|1+\zeta_{1}|^2}
\hspace{2mm} , \hspace{2mm}
\sigma \equiv \dfrac{i(\zeta_{1} - \bar{\zeta}_{1})}{|1+\zeta_{1}|^2}
\hspace{2mm} , \hspace{2mm}
\psi \equiv \dfrac{\zeta_{2}}{1+\zeta_{1}} \ ,
\end{equation}
and in terms of which the kinetic terms boil down to \cite{Bobev:2010ib,Halmagyi:2011xh}
\begin{equation}
\label{LagQK_2C}
\begin{array}{lll}
e^{-1} \mathcal{L}^{\textrm{QK}}_{\textrm{kin}}  &=& - \,\,\,\, \dfrac{(\partial \zeta_{1})(\partial \bar{\zeta}_{1}) + (\partial \zeta_{2})(\partial \bar{\zeta}_{2}) }{1 - |\zeta_{1}|^2 - |\zeta_{2}|^2 } \\[4mm]
&-& \dfrac{( \zeta_{1} \, \partial \bar{\zeta}_{1} + \zeta_{2} \, \partial \bar{\zeta}_{2})  ( \bar{\zeta}_{1} \, \partial \zeta_{1} + \bar{\zeta}_{2} \, \partial \zeta_{2}) }{\left( \, 1 - |\zeta_{1}|^2 - |\zeta_{2}|^2 \, \right)^2 } \ .
\end{array}
\end{equation}
The expressions (\ref{LagSK}), (\ref{LagQK_1C}) and (\ref{LagQK_2C}) often appear in the $\mathcal{N}=2$ literature (\textit{e.g.} see \cite{Bobev:2010ib,Halmagyi:2011xh}). Here we have shown their connection to the original scalars in (\ref{coset_SU3}) associated to the $\textrm{E}_{7(7)}$ generators displayed in (\ref{Gener_SU3}).

In the spirit of \cite{Bobev:2010ib}, and when restricted to the SU(3) invariant sector of the theories, the $\textrm{CSO}_{c}$ gaugings induce a scalar potential that can be derived from a ``superpotential"  $\mathcal{W}$. This superpotential takes the form \footnote{Notice its similarity with a central charge $Z = e^{K/2}  (g \mathcal{W}_{0} + i g c  \mathcal{W}_{\infty})$ in the flux vacua attractor mechanism of \cite{Kallosh:2005ax} with $K$ being the K\"ahler potential of a SK manifold $[\textrm{SU}(1,1)/\textrm{U}(1)] \times [\textrm{SU}(1,1)/\textrm{U}(1)]$ in the unit-disk parameterisation.}
\begin{equation}
\label{WN2}
\mathcal{W} =    (1-|z|^{2})^{-\frac{3}{2}} \, (1-|\zeta_{12}|^{2})^{-2} \,  (g \, \mathcal{W}_{0} + i \, g c \, \mathcal{W}_{\infty})  \ ,
\end{equation}
where $\,\mathcal{W}_{0}\,$ and $\,\mathcal{W}_{\infty}\,$ are  functions of the complex scalar $z$ in (\ref{SK_new variables}) and also of the scalars $(\zeta_{1},\zeta_{2})$ in (\ref{QK_new variables_2}). Computing the gauge covariant derivatives $\,D_{\mu} \mathcal{M}\,$ for the different gaugings using (\ref{D_cov_SO8}), (\ref{D_cov_SO7}) and (\ref{D_cov_SO6}), one finds that  $(\zeta_{1},\zeta_{2})$ actually enter the (super)potential through a \textit{gauge invariant} combination $\,\zeta_{12}\,$ satisfying $\,D_{\mu} \zeta_{12} = \partial_{\mu} \zeta_{12}$. For the SO(8)$_{c}$ and SO(6,2)$_{c}$ gaugings, the form of $\,\zeta_{12}\,$ is given by \cite{Bobev:2010ib}
\begin{equation}
\label{zeta_12_Def_sigma}
\zeta_{12}(\lambda,\sigma,|\psi|) =  \dfrac{|\zeta_{1}| + i \, |\zeta_{2}|}{1+\sqrt{1-|\zeta_{1}|^2 - |\zeta_{2}|^2}} \ .
\end{equation}
However, only when evaluated at $\sigma=0$, the combination in (\ref{zeta_12_Def_sigma}) proves to be gauge invariant also for the ISO(7)$_{c}$, ISO(6,1)$_{c}$ and SO(7,1)$_{c}$ gaugings. This is
\begin{equation}
\label{zeta_12_Def}
\zeta_{12} \equiv \zeta_{12}(\lambda,\sigma,|\psi|) \big|_{\sigma=0} = \frac{\Phi_{2}-i}{\Phi_{2}+i}   \ ,
\end{equation}
with $\,\Phi_{2} \equiv  -|\psi| + i \, \sqrt{\lambda}\,$. Setting $\,\sigma=0\,$ is compatible with the extremum condition $\,\partial_{\sigma} V \big|_{\sigma=0}=0\,$ for all the gaugings. In fact, due to the parameterisation in (\ref{Gener_SU3}) and (\ref{coset_SU3}), the scalar $\sigma$ does not enter the potential for the non-semisimple gaugings and does it quadratically (lowest order) for the semisimple ones. $V$ is also independent of $\textrm{Arg}(\psi)$ for all the gaugings \footnote{Alternatively, it is possible to mod out the theory by a $\mathbb{Z}_{2}$ element \cite{Borghese:2013dja} truncating away the $(a,\zeta)$ fields in (\ref{coset_SU3}) associated to the $(g_{2},g_{5})$ generators in (\ref{Gener_SU3}). This amounts to set $\sigma=\textrm{Arg}(\psi)=0$ in (\ref{QK_new variables_1}) and, additionally, the SU(3)-invariant vector fields are also projected out.}. Fixing $\sigma=\textrm{Arg}(\psi)=0$ in (\ref{LagQK_1C}) allows for a rewriting \cite{Bobev:2010ib}
\begin{equation}
\label{LagQK_3C}
e^{-1} \mathcal{L}^{\textrm{QK}}_{\textrm{kin}} =  4 \, \dfrac{\partial \Phi_{2} \,  \partial \bar{\Phi}_{2} }{( \Phi_{2} -\bar{\Phi}_{2})^2}   =  - 4 \, \dfrac{\partial \zeta_{12} \,  \partial \bar{\zeta}_{12}}{( 1- |\zeta_{12}|^2)^2}   \ ,
\end{equation}
making manifest the SK submanifold $\textrm{SU}(1,1)/\textrm{U}(1)\subset \textrm{SU}(2,1)/\textrm{U}(2)$ spanned by $\zeta_{12}$.

The superpotential $\,\mathcal{W}\,$ in (\ref{WN2}) is totally specified by $\mathcal{W}_{0}\,$ if $\,c=0$. In this limit, superpotentials were known for the electric SO(8)$_{c=0}$ theory \cite{Ahn:2000mf,Ahn:2009as,Bobev:2010ib} as well as for the SO(7,1)$_{c=0}$, ISO(7)$_{c=0}$ and SO(6,2)$_{c=0}$ ones \cite{Ahn:2000mf,Ahn:2001by,Ahn:2002qga,Ahn:2009as}. In the complementary limit  $c \rightarrow \infty$,  it is the function $\mathcal{W}_{\infty}$ that dominates. A derivation of the functions $\mathcal{W}_{0}$ and $\mathcal{W}_{\infty}$ gives the following results:

\begin{itemize}[leftmargin=4mm]

\item[$\circ$]  \textit{SO(8)$_{c}$ gaugings \cite{Borghese:2012zs,Guarino:2013gsa}}
\begin{equation}
\label{W01_SO8}
\begin{array}{llll}
\mathcal{W}_{0} & = & (1+z^{3}) \, (1+\zeta_{12}^4) + 6 \, z  \, (1+z)  \, \zeta_{12}^2  & ,\\[2mm]
\mathcal{W}_{\infty} & = & (1-z^{3})  \, (1+\zeta_{12}^4) - 6 \, z  \, (1-z)  \, \zeta_{12}^2 & ,
\end{array}
\end{equation}

\item[$\circ$]  \textit{SO(7,1)$_{c}$ gaugings}
\begin{equation}
\label{W01_SO71}
\begin{array}{llll}
\mathcal{W}_{0} & = & \tfrac{3}{4}   \,  (1 - \zeta _{12}^2 )^{2} \, (1-z^2) \, (1-z)  \\[2mm]
&-&  (1+z)^3  \, \zeta _{12}\,  (1 + \zeta _{12}^2) & ,\\[3mm]
\mathcal{W}_{\infty} & = & \tfrac{3}{4}   \,  (1 - \zeta _{12}^2 )^{2} \, (1-z^2) \, (1+z) \\[2mm]
&+&  (1-z)^3  \, \zeta _{12}\,  (1 + \zeta _{12}^2) & ,
\end{array}
\end{equation}

\item[$\circ$]  \textit{ISO(7)$_{c}$ gaugings \cite{Guarino:2015qaa}}
\begin{equation}
\label{W01_ISO7}
\begin{array}{llll}
\mathcal{W}_{0} & = &  \tfrac{7}{8} \,  (1- \zeta_{12})^{4} \, (1+z)^{3} \\[2mm]
&+& 3 \, (\zeta_{12}-z) \, (1+z) \,(1-\zeta_{12})^2 \,(1- z \, \zeta_{12})  & ,\\[3mm]
\mathcal{W}_{\infty} & = & \tfrac{1}{8} \,  (1-\zeta_{12})^{4} \, (1-z)^{3} & ,
\end{array}
\end{equation}

\item[$\circ$]  \textit{SO(6,2)$_{c}$ gaugings}
\begin{equation}
\label{W01_SO62}
\begin{array}{llll}
\mathcal{W}_{0} & = & \tfrac{1}{2} \,  (1+z) \, (1+z^2) \, (1-6 \, \zeta _{12}^2+\zeta _{12}^4) \\[2mm]
&-&  2  \, z \,  (1+z) \, (1+\zeta _{12}^4)  & ,\\[3mm]
\mathcal{W}_{\infty} & = & \tfrac{1}{2} \,  (1-z) \, (1+z^2) \, (1-6 \, \zeta _{12}^2+\zeta _{12}^4) \\[2mm]
&+& 2  \, z \,  (1-z) \, (1+\zeta _{12}^4)  & ,
\end{array}
\end{equation}

\item[$\circ$]  \textit{ISO(6,1)$_{c}$ gaugings}
\begin{equation}
\label{W01_ISO61}
\begin{array}{llll}
\mathcal{W}_{0} & = &  \tfrac{1}{8} \,  (1- \zeta_{12})^{2} \, (1+z) \\[2mm]
&& \left [ \, (1+z^2)  \big( 5 \, (1+\zeta_{12}^2) +14 \, \zeta_{12}   \big) \right. & \\[2mm]
&& \left. -\, 2 \, z \, \big( 7 \, (1+\zeta_{12}^2) +10 \, \zeta_{12}   \big) \right] & ,\\[3mm]
\mathcal{W}_{\infty} & = & \tfrac{1}{8} \,  (1-\zeta_{12})^{4} \, (1-z)^{3} & .
\end{array}
\end{equation}

\end{itemize}
The functions $\,\mathcal{W}_{0}\,$ and $\,\mathcal{W}_{\infty}\,$ in (\ref{W01_SO8})-(\ref{W01_ISO61}) completely specify the  superpotential (\ref{WN2}) for the set of $\textrm{CSO}_{c}$ gaugings compatible with an SU(3) truncation of maximal supergravity. For the semisimple gaugings, $\,\mathcal{W}_{0}\,$ and $\mathcal{W}_{\infty}\,$ are related to each other by $(z,\zeta_{12}) \leftrightarrow (-z,-\zeta_{12})\,$, \mbox{rendering} the two limits $\,c=0\,$ and $\,c=\infty\,$ equivalent. This property no longer holds for the non-semisimple \mbox{gaugings} which, however, turn out to have the same $\mathcal{W}_{\infty}\,$. We will come back to these two features in the next \mbox{section.}

The scalar potential $V$ is then obtained from $\,\mathcal{W}\,$ via the formula \cite{Bobev:2010ib}
\begin{equation}
\label{VN2}
\begin{array}{llll}
V &=& 2  \left[  \dfrac{4}{3} \,  (1-|z|^{2})^{2} \, \left| \dfrac{\partial |\mathcal{W}|}{\partial z} \right|^{2} \right.  & \\[2mm]
   &+&  \left.  (1-|\zeta_{12}|^{2})^{2} \, \left| \dfrac{\partial |\mathcal{W}|}{\partial \zeta_{12}} \right|^{2}   -3 \, |\mathcal{W}|^{2} \right] \ .
\end{array}
\end{equation}
In addition to the superpotential $\,\mathcal{W}(z,\zeta_{12})\,$ in (\ref{WN2}), there is a companion one $\,\widetilde{\mathcal{W}}(z,\zeta_{12})=\mathcal{W}(z,\bar{\zeta}_{12})\,$ from which the same scalar potential in (\ref{VN2}) follows \cite{Bobev:2010ib,Borghese:2012zs}. \mbox{Critical} points of $\,V\,$ preserving $\,\mathcal{N}=2\,$ and $\,\mathcal{N}=1\,$ supersymmetry exist. The former satisfy $\,\partial|\mathcal{W}|=0\,$ and $\,\partial|\mathcal{\widetilde{W}}|=0\,$ simultaneously. In contrast, those preserving $\,\mathcal{N}=1\,$ supersymmetry satisfy either one or the other condition.  For all the gaugings in (\ref{W01_SO8})--(\ref{W01_ISO61}), the potential (\ref{VN2}) matches the one obtained from (\ref{V_general}) using the embedding tensor formalism.

\section{$\mathcal{N}=1\,$ superpotentials}
\label{Sec:N1}

Let us now concentrate on a different truncation based on a ${\textrm{G}_{0}=\mathbb{Z}_{2} \times \textrm{SO}(3)}$ invariant sector \cite{Dibitetto:2012ia} of the maximal supergravity multiplet. The bosonic field content of this truncation consists of the metric field and six real scalars. The truncation works as follows: the $\mathbb{Z}_{2}$ factor truncates $\mathcal{N}=8$ supergravity to $\mathcal{N}=4$ supergravity coupled to six vector multiplets \cite{Dibitetto:2011eu}, whereas the additional $\textrm{SO}(3)$ factor further truncates to $\mathcal{N}=1$ supergravity coupled to three chiral multiplets and no vector multiplets \cite{Dibitetto:2011gm}.

In addition to the gaugings of the previous section, there are SO(5,3)$_{c}$, SO(4,4)$_{c}$ and ISO(4,3)$_{c}$ gaugings compatible with the ${\textrm{G}_{0}=\mathbb{Z}_{2} \times \textrm{SO}(3)}$ truncation. The SO(3) is located one level lower in the chain of embeddings (\ref{N=2_chain}), namely,
\begin{equation}
\label{N=1_chain}
\begin{array}{lcl}
& \textrm{SU}(3) & \\[1mm]
\textrm{SO}(8) \supset  \textrm{SO}(7) \supset  \textrm{G}_{2} \supset   
&  \textrm{or}& \supset  \textrm{SO}(3)  \ .  \\[1.5mm]
& \textrm{SO}(4) & 
\end{array}
\end{equation} 
Under ${\textrm{G}_{0}=\mathbb{Z}_{2} \times \textrm{SO}(3)}$, the $\textbf{8}$ gravitini of the maximal theory decompose as $\textbf{8} \rightarrow \textbf{1}_{(+)} + \textbf{1}_{(-)} + \textbf{3}_{(+)} + \textbf{3}_{(-)}$ where the $_{(\pm)}$ subscript denotes the $\mathbb{Z}_{2}$-parity of the corresponding SO(3) representation. As a result, there is one $\mathbb{Z}_{2}$-even singlet $\textbf{1}_{(+)}$ responsible for the $\mathcal{N}=1$ supersymmetry of the truncation. Notice that the invariant metric for the ISO(5,2)$_{c}$ gaugings in (\ref{eta_SO52}) is simply not compatible with the above decomposition.

The six real scalars in the truncation are this time associated to the following E$_{7(7)}$ generators in the SL(8) basis \cite{Guarino:2015qaa}
\begin{equation}
\label{Gener_Z2xSO3}
\begin{array}{cclc}
g_{1} &=&  \phantom{-}  {t_{3}}^{3} + {t_{5}}^{5} + {t_{7}}^{7}  +  {t_{2}}^{2} + {t_{4}}^{4} + {t_{6}}^{6} - 3 \, (  {t_{1}}^{1} +  {t_{8}}^{8} )& , \\[2mm]

g_{2} &=&  {t_{1}}^{1} + {t_{3}}^{3} + {t_{5}}^{5} + {t_{7}}^{7} - {t_{2}}^{2} - {t_{4}}^{4} - {t_{6}}^{6} - {t_{8}}^{8} & , \\[2mm]

g_{3} &=&   -  {t_{3}}^{3} - {t_{5}}^{5} - {t_{7}}^{7}  +  {t_{2}}^{2} + {t_{4}}^{4} + {t_{6}}^{6} + 3 \,  ({t_{1}}^{1} -  {t_{8}}^{8})  & , \\[2mm]

g_{4} &=&  t_{1238} + t_{1458} + t_{1678}  &  , \\[2mm]

g_{5} &=&   t_{8246}  & ,  \\[2mm]

g_{6} &=&   t_{2578} + t_{4738} + t_{6358}  & ,

\end{array}
\end{equation}
which can be used to build the coset representative ${\mathcal{V}=\mathcal{V}_{1} \times \mathcal{V}_{2} \times \mathcal{V}_{3}}$ upon the exponentiations
\begin{equation}
\begin{array}{llrr}
\mathcal{V}_{1} &=& e^{-12 \, \chi_{1} \, g_{4}} \, e^{\frac{1}{4} \, \varphi_{1} \, g_{1}} & , \\[2mm]
\mathcal{V}_{2} &=& e^{-12 \, \chi_{2} \, g_{5}} \, e^{\frac{1}{4} \, \varphi_{2} \, g_{2}} & , \\[2mm]
\mathcal{V}_{3} &=& e^{-12 \, \chi_{3} \, g_{6}} \, e^{\frac{1}{4} \, \varphi_{3} \, g_{3}} & .
\end{array}
\end{equation}
The coset representative $\mathcal{V}$ determines the scalar-dependent matrix $\,\mathcal{M}=\mathcal{V} \, \mathcal{V}^{t}\,$ and, after plugging into (\ref{L_scalars}), one obtains the kinetic terms
\begin{equation}
\label{LKin_SO3}
\begin{array}{lll}
e^{-1}  \mathcal{L}_{\textrm{kin}} &=& - \frac{3}{4} \left[ (\partial \varphi_{1} )^2 + e^{2 \varphi_{1}} \, (\partial \chi_{1} )^2\right]  \\[3mm]
&-&   \frac{1}{4} \left[ (\partial \varphi_{2} )^2 + e^{2 \varphi_{2}} \, (\partial \chi_{2} )^2\right] \\[3mm]
&-&   \frac{3}{4} \left[ (\partial \varphi_{3} )^2 + e^{2 \varphi_{3}} \, (\partial \chi_{3} )^2\right] \ ,
\end{array}
\end{equation}
and again a lengthy expression for the scalar potential $V$ that depends on the specific gauging. Analogously to the previous section, we will re-express the resulting Lagrangian in an $\mathcal{N}=1$ fashion. For this purpose, we first introduce three complex fields
\begin{equation}
\Phi_{I} = - \chi_{I} + i \, e^{-\varphi_{I}}
\hspace{5mm} \textrm{ with } \hspace{5mm} I=1,2,3 \ ,
\end{equation}
which span a K\"ahler manifold $\mathcal{M}_{\textrm{scal}}=[\textrm{SU}(1,1)/\textrm{U}(1)]^{3}$ specified by the  K\"ahler potential
\begin{equation}
\label{K_N1}
K = \sum_{I=1}^{3} - n_{I} \, \log[-i \, (\Phi_{I}-\bar{\Phi}_{I})]    \ ,
\end{equation}
with $\,(n_{1},n_{2},n_{3})=(3,1,3)\,$. In terms of (\ref{K_N1}), the kinetic terms in (\ref{LKin_SO3}) can be rewritten as
\begin{equation}
\label{LagKinSL2^3}
e^{-1} \mathcal{L}_{\textrm{kin}} =  -\sum_{I=1}^{3} K_{\Phi_{I} \bar{\Phi}_{I}} \, \partial \Phi_{I} \,  \partial \bar{\Phi}_{I}    =   \sum_{I=1}^{3} n_{I} \, \frac{\partial \Phi_{I} \,  \partial \bar{\Phi}_{I} }{( \Phi_{I} -\bar{\Phi}_{I})^2}   \  ,
\end{equation}
where ${K_{\Phi_{I} \bar{\Phi}_{I}}=\partial_{\Phi_{I}}\partial_{\bar{\Phi}_{I}} K}$ is the K\"ahler metric. The interaction between the complex scalars $\Phi_{I}$ is encoded into an $\mathcal{N}=1$ holomorphic superpotential $W$ of the form
\begin{equation}
\label{W_N1}
W = g \, W_{0} + gc \, W_{\infty} \ ,
\end{equation}
where the functions $W_{0}$ and $W_{\infty}$ are given by
\begin{equation}
\label{WN1_all}
\begin{array}{lll}
W_{0} & = &     p_{1} \Phi_{1}^3 + 3 \, p_{2} \,  \Phi_{1} \Phi_{3}^2 + 3 \, p_{3} \, \Phi_{1}  \Phi_{2} \Phi_{3}+ p_{4} \, \Phi_{1}^3 \Phi_{2}  \Phi_{3}^3  \ ,  \\[2mm]
W_{\infty} & = &   q_{1} \,  \Phi_{2}  \Phi_{3}^3  + 3 \, q_{2} \, \Phi_{1}^2 \Phi_{2} \Phi_{3} + 3 \, q_{3}  \,\Phi_{1}^2  \Phi_{3}^2 + q_{4}   \  .
\end{array}
\end{equation}
The structure of monomials in $\,W_{0}\,$ and $\,W_{\infty}\,$ is such that, up to an overall sign, they map into each other upon a modular transformation $\,\Phi_{I}  \rightarrow -\Phi_{I}^{-1}\,$ followed by an exchange $\,p_{i} \leftrightarrow q_{i}\,$ \footnote{Since the monomials in $\,W_{0}\,$ and $\,W_{\infty}\,$ are related by the inversion $\,\Phi_{I}  \rightarrow -\Phi_{I}^{-1}\,$, their simultaneous presence in (\ref{W_N1}) when $c \neq 0$ democratises the set of positive and negative dilaton weight-vectors in the scalar potential for semisimple gaugings. This somehow resembles the double coset construction of \cite{Cremmer:1997ct,Cremmer:1998px}.}. The coefficients $\,p_{1,2,3,4}\,$ and $q_{1,2,3,4}\,$ in (\ref{WN1_all}) are displayed in Table~\ref{Table:N1+coeff} for all the gaugings compatible with ${\textrm{G}_{0}=\mathbb{Z}_{2} \times \textrm{SO}(3)}$. \mbox{Semisimple} gaugings come out with $\,p_{i} = q_{i} \neq 0\,$, $\,{\forall i=1,...,4}\,$, \mbox{rendering} the two limiting cases $\,c=0\,$ and $\,c=\infty\,$ equivalent. For a generic value of $\,c\,$, the superpotential (\ref{W_N1}) can be thought of as an inequivalent superposition of equivalent theories. In contrast, non-semisimple gaugings have $\,p_{i} \, q_{i} = 0\,$, $\forall i=1,...,4\,$, and this orthogonality between $\,W_{0}\,$ and $W_{\infty}\,$ makes the $\,c=0\,$ and $\,c=\infty\,$ cases no longer equivalent. Actually, all the non-semisimple gaugings in \mbox{Table~\ref{Table:N1+coeff}} become degenerated at the level of superpotentials in the $c \rightarrow \infty$ limit, namely, 
\begin{equation}
\label{WUniversal}
\displaystyle\lim_{c \rightarrow \infty} W_{\textrm{ISO}(p,q)_{c}} =2 \, g \, c \ .
\end{equation}
We will recall this ``universality" property later on when discussing possible higher-dimensional descriptions of the ISO$(p,q)_{c}$ gaugings. 

Using $K$ and $W$ in (\ref{K_N1}) and (\ref{W_N1}), the scalar potential follows from the standard $\mathcal{N}=1$ formula
\begin{equation}
\label{V_N=1}
V = e^{K} \left[ K^{\Phi_{I} \bar{\Phi}_{I}} (D_{\Phi_{I}} W) (D_{\bar{\Phi}_{I}} \bar{W}) - 3 \, W \, \bar{W} \right] \ ,
\end{equation}
where $K^{\Phi_{I} \bar{\Phi}_{I}}$ is the inverse of the K\"ahler metric in (\ref{LagKinSL2^3}) and $D_{\Phi_{I}} W=\partial_{\Phi_{I}} W + (\partial_{\Phi_{I}}K) W$ is the K\"ahler derivative. For all the gaugings in Table~\ref{Table:N1+coeff}, we have verified that the scalar potential (\ref{V_N=1}) exactly reproduces the one in (\ref{V_general}) using the embedding tensor formalism.

\begin{table}[t]
\renewcommand{\arraystretch}{1.5}
\begin{tabular}{!{\vrule width 1.5pt}c!{\vrule width 1pt}cccc!{\vrule width 1pt}cccc!{\vrule width 1.5pt}}
\noalign{\hrule height 1.5pt}
 \,\,\textsc{gauging} \,\,                     &  $\,\,p_{1}\,\,$ & $\,\,p_{2}\,\,$ & $\,\,p_{3}\,\,$ & $\,\,p_{4}\,\,$ & $\,\,q_{1}\,\,$ & $\,\,q_{2}\,\,$ & $\,\,q_{3}\,\,$ & $\,\,q_{4}\,\,$  \\
\noalign{\hrule height 1pt}
 $\textrm{SO}(8)_{c}$   &   $+2$ & $+2$ & $+2$ & $+2$ & $+2$ & $+2$ & $+2$ & $+2$  \\
 $\textrm{SO}(7,1)_{c}$   &   $+2$ & $+2$ & $+2$ & $-2$ & $+2$ & $+2$ & $+2$ & $-2$  \\
 $\textrm{ISO}(7)_{c}$   &   $+2$ & $+2$ & $+2$ & $0$ & $0$ & $0$ & $0$ & $+2$  \\
 $\textrm{SO}(6,2)_{c}$   &   $-2$ & $+2$ & $+2$ & $-2$ & $-2$ & $+2$ & $+2$ & $-2$  \\
 $\textrm{ISO}(6,1)_{c}$   &   $-2$ & $+2$ & $+2$ & $0$ & $0$ & $0$ & $0$ & $+2$  \\
\noalign{\hrule height 1pt}
 $\textrm{SO}(5,3)_{c}$   &   $+2$ & $+2$ & $-2$ & $+2$ & $+2$ & $+2$ & $-2$ & $+2$  \\
 $\textrm{SO}(4,4)_{c}$   &   $+2$ & $+2$ & $-2$ & $-2$ & $+2$ & $+2$ & $-2$ & $-2$  \\
 $\textrm{ISO}(4,3)_{c}$   &   $+2$ & $+2$ & $-2$ & $0$ & $0$ & $0$ & $0$ & $+2$  \\
 \noalign{\hrule height 1.5pt}
\end{tabular}
\caption{List of coefficients determining the $W_{0}$ and $W_{\infty}$ functions in (\ref{WN1_all}) for the set of CSO$_{c}$ gaugings compatible with ${\textrm{G}_{0}=\mathbb{Z}_{2} \times \textrm{SO}(3)}$.}
\label{Table:N1+coeff}
\end{table}

Taking a second look at the scalars in (\ref{Gener_SU3}) and (\ref{Gener_Z2xSO3}), there exists an overlapping between the $\mathcal{N}=1$ truncation based on ${\textrm{G}_{0}=\mathbb{Z}_{2} \times \textrm{SO}(3)}$ and the $\mathcal{N}=2$ truncation based on ${\textrm{G}_{0}=\textrm{SU}(3)}$ discussed in the previous section. This fact was already pointed out in \cite{Guarino:2013gsa} for the case of the SO(8)$_{c}$ gaugings and actually extends to all the CSO$_{c}$ gaugings in the upper block of Table~\ref{Table:N1+coeff}. The overlap between the two truncations proves an $\mathcal{N}=1$ supergravity coupled this time to two chiral multiplets and is realised by firstly identifying
\begin{equation}
\label{Phi_ident_SO4}
\Phi_{2}=\Phi_{3} \ ,
\end{equation}
and then applying the field redefinitions in (\ref{SK_new variables}) and (\ref{zeta_12_Def}) to the variables $z$ and $\zeta_{12}$, respectively. The modular transformation $\Phi_{I} \rightarrow -\Phi_{I}^{-1}$ then translates into the $(z , \zeta_{12}) \rightarrow (-z , -\zeta_{12})$ transformation discussed in the previous section.

Next comes the $\textrm{G}_{0}=\textrm{SO}(4)\sim \textrm{SO}(3) \times \textrm{SO}(3)$ invariant sector in (\ref{N=1_chain}). The specific embedding we consider here coincides with the one in \cite{Gallerati:2014xra} and is compatible with SO(8)$_{c}$, SO(7,1)$_{c}$, ISO(7)$_{c}$, SO(5,3)$_{c}$, SO(4,4)$_{c}$ and ISO(4,3)$_{c}$ gaugings. It is recovered from the ${\textrm{G}_{0}=\mathbb{Z}_{2} \times \textrm{SO}(3)}$ sector by identifying
\begin{equation}
\label{Phi1=Phi3}
\Phi_{1}=\Phi_{3} \ .
\end{equation}
This truncation produces the gravitini decomposition $\textbf{8} \rightarrow (\textbf{1},\textbf{1}) + (\textbf{3},\textbf{1}) + (\textbf{2},\textbf{2})$ and also corresponds to $\mathcal{N}=1$ supergravity coupled to two chiral multiplets \footnote{The set of CSO$_{c}$ gaugings in Table~\ref{Table:N1+coeff} may accommodate different $\textrm{G}_{0}=\textrm{SO}(4)$ invariant sectors. For example, some non-supersymmetric $\textrm{SO}(4)$ truncations of the SO(4,4)$_{c}$ and SO(6,2)$_{c}$ gaugings specified by different decompositions of the $\textbf{8}$ gravitini were investigated in \cite{Dall'Agata:2012sx,Borghese:2013dja}.}. For the ISO(7)$_{c}$ gaugings, this sector has been thoroughly investigated in \cite{Guarino:2015qaa} (see \cite{Pang:2015mra} for the SO(8)$_{c}$ gaugings) and found to contain the novel critical point of \cite{Gallerati:2014xra} preserving $\mathcal{N}=3$ supersymmetry within the full theory. Here we have verified that similar $\mathcal{N}=3$ critical points also exist for the SO(8)$_{c}$ and the SO(7,1)$_{c}$ gaugings, in agreement with \cite{Gallerati:2014xra}. In addition, the SO(7,1)$_{c}$ family of gaugings was found to include an $\,\mathcal{N}=4\,$ critical point preserving a different SO(4) subgroup \cite{Gallerati:2014xra}. This point corresponds to $\Phi_{1}=i \, c\,$ and $\,\Phi_{2}=-\bar{\Phi}_{3}=e^{i \, \pi/4}\,$ or $\,-\bar{\Phi}_{2}=\Phi_{3}=e^{i \, \pi/4}\,$. In the first case, only one out of the four supersymmetries preserved by the solution lies within the $\,\mathcal{N}=4\,$ theory obtained as $\,\mathcal{N}=8 \overset{\mathbb{Z}_{2}}{\rightarrow} \mathcal{N}=4\,$. In the second case, three out of the four supersymmetries belong to the $\mathcal{N}=4$ theory.

A truncation based on $\textrm{G}_{0}=\textrm{G}_{2}$ is compatible only with SO(8)$_{c}$, SO(7,1)$_{c}$ and ISO(7)$_{c}$ gaugings \cite{Borghese:2012qm}. The decomposition of the $\textbf{8}$ gravitini reads $\textbf{8} \rightarrow \textbf{1} + \textbf{7}$ and the truncation corresponds to $\mathcal{N}=1$ supergravity coupled to one chiral multiplet and no vectors. This sector is recovered upon the identification
\begin{equation}
\label{Phi123}
\Phi_{1}=\Phi_{2}=\Phi_{3} \ ,
\end{equation}
of the three chiral fields in the ${\textrm{G}_{0}=\mathbb{Z}_{2} \times \textrm{SO}(3)}$ sector. 
In addition to the identifications in (\ref{Phi_ident_SO4})--(\ref{Phi123}), there are equivalent ones obtained by discrete transformations. \mbox{Finally}, there also exist $\textrm{G}_{0}=\textrm{SO}(7)$ and $\textrm{G}_{0}=\textrm{SO}(6)$ invariant sectors but these produce non-supersymmetric truncations, so we are not considering them further in this note.

\section{(Non-)geometric string/M-theory backgrounds}
\label{Sec:String/M-theory}

Some of the $\mathcal{N}=1$ supergravities specified in \mbox{Table~\ref{Table:N1+coeff}} have appeared in the context of non-geometric flux compactifications on toroidal backgrounds \footnote{The dependence on the duality frame (type IIA/IIB, \mbox{Heterotic}, M-theory,..) when it comes to classify a given supergravity as a geometric/non-geometric toroidal background has been discussed in \cite{Shelton:2005cf,Aldazabal:2006up,Dibitetto:2011gm,Derendinger:2014wwa}.}. We will \mbox{consider} type IIB orientifolds of $\,T^{6}/(\mathbb{Z}_{2}\times\mathbb{Z}_{2})$ with O3/O7-planes yielding the so-called STU-models \cite{Shelton:2005cf,Aldazabal:2006up,Dibitetto:2011gm,Dibitetto:2012ia} in the isotropic (or plane-exchange-symmetric) limit \cite{Derendinger:2004jn}. The connection to the dilaton ($S$), K\"ahler ($T$) and complex structure ($U$) moduli in \cite{Dibitetto:2011gm,Dibitetto:2012ia} is given by 
\begin{equation}
\label{STU_identifications}
\Phi_{1} = -U^{-1}
\hspace{4mm} , \hspace{4mm}
\Phi_{2} = S
\hspace{4mm} , \hspace{4mm}
\Phi_{3} = T \ .
\end{equation}
Upon a modular transformation $\,U \rightarrow -U^{-1}\,$, one \mbox{obtains} standard \mbox{STU-models} specified by the K\"ahler potential 
\begin{equation}
\label{KahlerSTU}
\begin{array}{llll}
K &=& - \, 3 \, \log[-i \, (U-\bar{U})] & \\[2mm]
    & &  - \,3 \, \log[-i \, (T-\bar{T})] \, - \, \log[-i \, (S-\bar{S})]    & ,
\end{array}
\end{equation}
and the $\,\mathcal{N}=1\,$ superpotential (\ref{W_N1}) with
\begin{equation}
\label{WN1_all_STU}
\begin{array}{llll}
W_{0} & = &   -p_{1}  - 3 \, p_{2} \,  U^2 T^2 - 3 \, p_{3} \, U^2  S T- p_{4} \, S  T^3 & ,  \\[2mm]
W_{\infty} & = &   q_{1} \,  U^3 S  T^3  + 3 \, q_{2} \, U S T + 3 \, q_{3}  \, U  T^2 + q_{4} \, U^3   & .
\end{array}
\end{equation}
The mapping between the coefficients $\,(p_{i},q_{i})\,$ in (\ref{WN1_all_STU}) and the generalised type IIB fluxes in \cite{Dibitetto:2011gm,Dibitetto:2012ia} reads
\begin{equation}
\label{Flux_mapping}
\begin{array}{llllllll}
p_{1} \equiv F_3  \hspace{2mm}&  ,  & \hspace{2mm}p_{2} \equiv Q'  \hspace{2mm}& , & \hspace{2mm}p_{3} \equiv P  & , & \hspace{2mm}p_{4} \equiv H_{3}'  & , \\[2mm]
q_{1} \equiv H_{3}'  \hspace{2mm}&  ,  & \hspace{2mm}q_{2} \equiv P  \hspace{2mm}& , & \hspace{2mm}q_{3} \equiv Q'  & , & \hspace{2mm}q_{4} \equiv F_{3}  & ,
\end{array}
\end{equation} 
where $F_3$ is a Ramond-Ramond three-form flux and \mbox{$H_{3}'$ , $Q'$ and $P$} are (highly) non-geometric fluxes. As a result, these supergravities correspond to non-geometric type IIB toroidal backgrounds for \textit{any} value of the electric/magnetic parameter, including the $\,c=0\,$ case. 

However, when $c=0$, a geometric description in terms of M-theory reductions on non-compact ${\mathcal{H}^{p,q,r}=\mathcal{H}^{p,q} \times T^{r}}$ spaces, with $\,\mathcal{H}^{p,q}\,$ being a hyperboloid, is available for the CSO$(p,q,r)$ gaugings \cite{Hull:1988jw,Gibbons:2001wy,Baron:2014bya}. The observation that non-geometric toroidal backgrounds may still admit geometric descriptions as non-toroidal reductions has already been made in the literature, \textit{e.g.}, see appendix~A of \cite{Catino:2013ppa} or also \cite{Danielsson:2015tsa} for a more  recent discussion on non-geometric STU-models linked to compactifications of M-theory. In this regard, the K\"ahler potential in (\ref{KahlerSTU}) and the $W_{0}$ superpotential in (\ref{WN1_all_STU}) provide further examples of STU-models for the electric $\textrm{CSO}_{c=0}\,$ gaugings in Table~\ref{Table:N1+coeff} that connect to \mbox{M-theory} reductions on $\mathcal{H}^{p,q,r}$ spaces. 

Turning on $\,c\neq0\,$ generically causes the loss of a higher-dimensional interpretation of the CSO$_{c}$ maximal supergravities. The universal limit (\ref{WUniversal}) suggests a possible ten-dimensional description of the $\,\textrm{ISO}(p,q)_{c}\,$ gaugings in terms of \textit{massive} type IIA reductions on $\,\mathcal{H}^{p,q}\,$ spaces along the lines of \cite{Guarino:2015jca,Guarino:xxx2}. Taking ${c\rightarrow \infty}$, which is identified with taking a (infinitely) large \mbox{Romans} mass ${m=gc}\,$ in \cite{Guarino:2015jca}, was also linked to a regular reduction of massive type IIA on $T^{6}$ in \cite{Guarino:2015qaa}. This purely magnetic limit would hide the dependence of the $\textrm{ISO}(p,q)_{c}$ \mbox{maximal} supergravities on the $\,\mathcal{H}^{p,q}\,$ geometries clearly visible at $\,{c=0}\,$, resulting in the universal \mbox{superpotential} of (\ref{WUniversal}). For the STU-models in (\ref{WN1_all_STU}), this becomes ${\lim_{c \rightarrow \infty} W_{\textrm{ISO}(p,q)_{c}} =2 \, gc \, U^3}\,$ and agrees with the identification done in \cite{Derendinger:2004jn,Shelton:2005cf} \footnote{The ($S,T,U$) fields here were denoted $(S,U,\tau)$ in \cite{Shelton:2005cf} and $(S,U,T)$ in \cite{Derendinger:2004jn}.} between the $U^3$ coupling in the superpotential and the Romans mass parameter $m=gc\,$ in a type IIA incarnation of the flux models \footnote{See \mbox{appendix~A} of \cite{Guarino:2015qaa} for a detailed discussion on the non-geometric STU-model associated to the ISO(7)$_{c}$ gaugings and its geometric origin as a massive type IIA reduction on $\,S^6$ \cite{Guarino:2015jca,Guarino:xxx2}. Note also that, in a IIA picture \cite{Derendinger:2004jn}, the moduli $\,T\,$ and $\,U\,$ in (\ref{KahlerSTU})--(\ref{WN1_all_STU}) are identified as complex structure and K\"ahler moduli, respectively.}.

Finding string/M-theory candidates to describe the semisimple SO$(p,q)_{c}$ gaugings proves a more challenging task \footnote{For semisimple $\textrm{SO}(p,q)_{c}$ gaugings one could speculate on the limiting $\,c = \infty\,$ case as an M-theory reduction, equivalent to one at $c=0$, but on a ``dual'' $\mathcal{H}^{p,q}$ space. \mbox{In the context} of Generalised Geometry, refs~\cite{Lee:2014mla,Lee:2015xga} studied the SO(8)$_{c}$ gaugings at generic values of $\,c\,$ and showed that these cannot be realised as a compactification of a higher-dimensional theory that is locally geometrical. This is in line with the highly non-geometric fluxes in (\ref{Flux_mapping}) underlying the $\textrm{SO}(p,q)_{c}$ gaugings.}. Symplectic deformations of semisimple gaugings -- see (\ref{D_cov_SO8}), (\ref{D_cov_SO7}) and (\ref{D_cov_SO6}) -- involve magnetic vectors linked to compact ($com$) generators of $\,\textrm{G}\,$ and, consistently, two-form tensor fields in the modified electric field strengths \cite{deWit:2005ub,deWit:2007mt}. \mbox{Schematically}, ${\mathcal{H}^{com}_{\mu\nu}=\mathcal{F}^{com}_{\mu\nu} - \frac{1}{2} \, g c \, B_{\mu\nu}}\,$, where $\,\mathcal{F}^{com}_{\mu\nu}\,$ is the electric Yang-Mills field strength and $\,B_{\mu \nu}\,$ is a two-form field. The \mbox{inclusion} of tensor fields $B_{\mu\nu}$'s linked to compact generators of the U-duality group has recently been discussed in \cite{Bandos:2015ila}. Together with those which are dual to scalars in the $\textrm{E}_{7(7)}/\textrm{SU}(8)$ coset, the ``compact" tensor fields -- dubbed ``auxiliary notophs" in \cite{Ogievetsky:1967ij,Bandos:2015ila} --  are necessary ingredients in a superspace formulation of ungauged maximal supergravity and happen to be dual to fermion bilinears. In this sense, a gauged version of the superfield description of notophs, as well as the search for worldvolume actions, might shed light upon the microscopic origin, if any, of the SO$(p,q)_{c}$ maximal \mbox{supergravities.}

\section{Summary and discussion}
\label{Sec:Summary}

In this note we have provided a collection of superpotentials controlling the scalar dynamics in certain $\,\mathcal{N}=2\,$ and $\,\mathcal{N}=1\,$ supergravities in four dimensions. Despite their simplicity, these supergravities describe consistent truncations of the one-parameter family (with parameter $c$) of \mbox{electric/magnetic} SO(8)$_{c}$ gauged supergravities discovered in \cite{Dall'Agata:2012bb} as well as its generalisation to other CSO$(p,q,r)_{c}$ gaugings. We have studied two different truncations producing an Einstein-scalar Lagrangian of the form
\begin{equation}
\label{L_scalars_summary}
\mathcal{L}_{\textrm{E-s}} = \tfrac{1}{2} \, e \, R + \mathcal{L}_{\textrm{kin}} - e \,  V \ .
\end{equation}
Here is a summary of the main results:
\begin{itemize}[leftmargin=4mm]
\item[$i)$] The first truncation, see \cref{Sec:N2},  is based on a ${\textrm{G}_{0}=\textrm{SU}(3)}$ invariant sector and produces an $\mathcal{N}=2$ supergravity coupled to one vector multiplet (with complex scalar $z$) and one hypermultiplet (with complex scalars $\zeta_{1}$ and $\zeta_{2}$). $\mathcal{L}_{\textrm{kin}}$ consists of the two pieces (\ref{LagSK}) and (\ref{LagQK_3C}) and, in the spirit of \cite{Bobev:2010ib}, $V$ is obtained from the electric/magnetic superpotential $\mathcal{W}$ in (\ref{WN2}) using (\ref{VN2}). The different superpotentials associated to the different gaugings compatible with the truncation are listed in (\ref{W01_SO8})--(\ref{W01_ISO61}). 
\item[$ii)$] The second truncation, see \cref{Sec:N1},  is based on a ${\textrm{G}_{0}=\mathbb{Z}_{2} \times \textrm{SO}(3)}$ invariant sector and yields an $\mathcal{N}=1$ supergravity coupled to three chiral multiplets $\Phi_{1,2,3}$. The kinetic piece $\mathcal{L}_{\textrm{kin}}$ takes the form (\ref{LagKinSL2^3}) and the potential $V$ follows from the electric/magnetic superpotential $W$ in (\ref{W_N1}) using the standard formula (\ref{V_N=1}). The different superpotentials for the different gaugings compatible with the truncation are encoded in the coefficients displayed in Table~\ref{Table:N1+coeff}. 
\end{itemize}

Applications of the CSO$_{c}$ superpotentials presented in this note are immediately envisaged. The first one is the dedicated exploration and classification of critical points of the associated scalar potentials. The amount of both supersymmetric and non-supersymmetric critical points generically increases when $c\neq0$ \cite{DallAgata:2011aa,Borghese:2012qm,Dall'Agata:2012sx,Borghese:2013dja,Guarino:2015qaa}, especially for non-compact gaugings \cite{Hull:1984vg,Hull:1984ea,Hull:1984rt,Ahn:2001by,Ahn:2002qga}. Supersymmetric extrema play a central role in the construction of BPS domain-wall solutions which are conjectured to describe RG flows via the gauge/gravity correspondence. Such BPS domain-walls have been extensively studied within the $\textrm{G}_{0}=\textrm{SU}(3)$ invariant sector of the CSO$_{c=0}$ maximal supergravities \cite{Ahn:2000mf,Ahn:2001by,Ahn:2002qga} and, more recently, also of the family of SO(8)$_{c}$ gaugings \cite{Guarino:2013gsa,Tarrio:2013qga} using  the superpotential in (\ref{WN2}) and (\ref{W01_SO8}). Therefore, a second application of the CSO$_{c}$ superpotentials is the systematic study of BPS domain-walls within the $\textrm{G}_{0}=\textrm{SU}(3)$ and, even more general, within the $\textrm{G}_{0}=\mathbb{Z}_{2} \times \textrm{SO}(3)$ invariant sectors of the theories, as well as their dual RG flows. It would also be interesting to search for hairy black holes in these Einstein-scalar systems with a potential. Finally, the existence of a higher-dimensional description for the electric/magnetic families of CSO$_{c}$ maximal supergravities remains one of the essential questions to be answered. By looking at the  CSO$_{c}$ superpotentials, the electric/magnetic deformation interpolates between two equivalent theories at $c=0$ and $c=\infty$ for semisimple gaugings, whereas non-semisimple gaugings flow towards the universal superpotential (\ref{WUniversal}) in the $c \rightarrow \infty\,$ limit. This fact makes the massive type IIA reductions on $\mathcal{H}^{p,q}\,$ spaces natural scenarios where to investigate the higher-dimensional origin of the ISO$(p,q)_{c}$ maximal supergravities, just like the ISO$(7)_{c}$ gaugings \cite{Guarino:2015qaa} have recently been connected to such reductions on $\,\mathcal{H}^{7,0}=S^{6}\,$ in \cite{Guarino:2015jca,Guarino:xxx2}. In contrast, the semisimple SO$(p,q)_{c}$ gaugings remain elusive and alternative approaches, as the one based on a (gauged version of) superspace formulation of maximal supergravity \cite{Bandos:2015ila} or those of Exceptional Generalised Geometry \cite{Hull:2007zu,Pacheco:2008ps} and  Exceptional Field Theory \cite{Hohm:2013pua,Hohm:2013uia}, are at this time under \mbox{investigation}. We hope to come back to these issues in the near future.
\\

\noindent{\bf Acknowledgements:} We are especially grateful to Bernard de Wit and NIKHEF for their support and flexibility during the elaboration of this note. We want to thank Gianluca Inverso for useful explanations of his work \cite{Dall'Agata:2014ita} and Oscar Varela for stimulating conversations and collaboration in related projects. Finally, we also thank Mario Trigiante for a correspondence on his work \cite{Gallerati:2014xra}. The work of AG is supported by the ERC Advanced Grant no. 246974, ``Supersymmetry: a window to non-perturbative physics".

\section*{Appendix: SL(8) basis of E$_{7(7)}$ generators}
\label{sec:appendix}

Let us introduce a fundamental SL(8) index $A=1,...8$. In the SL(8) basis, the E$_{7(7)}$ generators $t_{\alpha=1,...,133}$ have a decomposition $\,\textbf{133} \rightarrow \textbf{63} + \textbf{70}\,$. These are the \textbf{63} generators ${t_{A}}^{B}$ of SL(8), with ${t_{A}}^{A}=0$, together with \textbf{70} generators $t_{ABCD}=t_{[ABCD]}$. The fundamental representation of E$_{7(7)}$ decomposes as $\,\textbf{56} \rightarrow \textbf{28} + \textbf{28'}\,$, what translates into an index splitting $\,_\mathbb{M} \rightarrow _{[AB]} \oplus ^{[AB]}$. The entries of the $\,56 \times 56\,$ matrices $\,{[t_{\alpha}]_{\mathbb{M}}}^{\mathbb{N}\,}$ are given by
\begin{equation}
\label{63Gener}
\begin{array}{llll}
{[{t_{A}}^{B}]_{[CD]}}^{[EF]} &=& 4 \, \left( \delta_{[C}^{B} \, \delta_{D]A}^{EF} + \frac{1}{8} \, \delta_{A}^{B} \, \delta_{CD}^{EF}  \right) & ,\\[2mm]
{[{t_{A}}^{B}]^{[EF]}}_{[CD]} &=& - {[{t_{A}}^{B}]_{[CD]}}^{[EF]} & ,
\end{array}
\end{equation}
for the SL(8) generators ${t_{A}}^{B}$ and by
\begin{equation}
\label{70Gener}
\begin{array}{llll}
[t_{ABCD}]_{[EF][GH]} &=& \frac{2}{4!} \, \epsilon_{ABCDEFGH} & , \\[2mm]
[t_{ABCD}]^{[EF][GH]} &=& 2 \, \delta_{ABCD}^{EFGH} & ,
\end{array}
\end{equation}
for the generators $t_{ABCD}$ extending to E$_{7(7)}$.


\end{document}